\documentclass[aps,pra,preprint,longbibliography,nofootinbib]{revtex4-2}

\makeatletter
\renewcommand{\p@subsection}{}
\renewcommand{\p@subsubsection}{}
\makeatother

\usepackage[utf8]{inputenc}
\usepackage{lmodern} 
\usepackage{textcomp}
\usepackage{geometry}
\usepackage{setspace}
\usepackage[pdftex, plainpages=false]{hyperref}
\usepackage{amsmath, amssymb, amsthm, mathtools}
\usepackage{enumitem}
\usepackage{float}
\usepackage[titletoc]{appendix}
\usepackage{tcolorbox}
\usepackage{tikz}
\usepackage[]{xcolor}

\usepackage{physics}

\newcommand{\I}[0]{\mathcal{I}}

\renewcommand{\L}[0]{\mathcal{L}}

\newcommand{\Q}[0]{\mathcal{Q}}

\renewcommand{\dd}{\mathrm{d}}

\definecolor{darkgreen}{rgb}{0.0, 0.5, 0.0}
\usepackage{subcaption}

\newtheorem{definition}{Definition}

\begin{document}
\title{Measuring inequality and social stratification with Lorenz curvature}
\author{Antti Hippel\"ainen}
\email{antti.hippelainen@helsinki.fi}
\affiliation{Department of Physics and Helsinki Institute of Physics, P.O. Box 64, FI-00014, University of Helsinki, Finland}

\begin{abstract}
We construct a continuous family of inequality and social stratification indices based on the curvature of the Lorenz curve. We study the inequality axioms of the family and find that they are satisfied only with the so-called stratification-aversion parameter $\alpha$ set to 0 -- in all other cases, these constraints cannot be satisfied in a strict sense. With $\alpha = 0$, the index has a very simple closed form and its value can be easily approximated. We study the values of the index on World Bank inequality data and see how the rankings across a wide selection of countries change as $\alpha$ is varied. Spearman and Kendall rank-correlation matrices with other common indices are also computed and analyzed. We find the curvature index to deviate consistently from the comparison indices, albeit less so for consumption- than income-based countries.
\end{abstract}

\maketitle
\clearpage

\tableofcontents

\section{Introduction}
\label{sec:introduction}

Measurement of economic inequality is an enduring challenge of quantitative economics \cite{OxfordHandbook}. This challenge is compounded by the multifaceted nature of inequality, ranging from a simple lack of monetary resources to social exclusion and societal instability.

Traditionally, study of economic inequality has been led by the study of the Lorenz curve \cite{Lorenz1905}. For each population fraction, the Lorenz curve represents the smallest amount of an economic resource (\textit{e.g.} income, wealth, consumption) controlled by the given fraction. 

Inequality indices like the Gini coefficient \cite{Gini1912book, Gini1921paper, cowell2011measuring} aggregate the deviation of a given population from perfect equality, condensing inequality into a single scalar value. Thus, very differently structured inequality may result in the same value of inequality, and while effective at summarizing, such indices are not sensitive to at what population fraction inequality is most prevalent. 

With the Lorenz curve $\L(x) = x$ describing perfect equality, intuition suggests that inequality is connected to the deviation of the Lorenz curve from the line of perfect equality. We propose studying this deviation with the curvature of the Lorenz curve. A region of high curvature corresponds to low income-distribution density, giving a natural signature of a gap between economic classes. Averaging over a local property such as curvature also allows one to define global, single scalar-indices.

A somewhat reminiscent index was constructed by Amato and Kakwani \cite{Amato1968, Kakwani1980, Arnold2012}, which computes the length of the Lorenz curve, normalizing with respect to the length of the line of perfect equality. The original construction by Atkinson \cite{Atkinson1970} is also reminiscent with its $L^p$- or $\ell^p$-norm structure, and we give a comparison of these two indices with the curvature-based one.

The curvature-based measures we propose measure at each point of the Lorenz curve how tight the bending of the curve is, at the same time capturing the intensity of stratification. To the best of our knowledge, no prior inequality index has been constructed from the curvature of the Lorenz curve.  

The rest of the work is organized as follows: In Section \ref{sec:curvaturebasedinequality} we establish the mathematical framework, first laying out the basic principles of Lorenz curves and curvature. We combine different possible choices under a one-parameter family of indices. In Section \ref{sec:axiomaticproperties} we study the fulfillment of inequality index axioms of this index family, highlighting the so-called stratification-aversion parameter $\alpha$ and its effect. In Section \ref{sec:globalprospectsandcomparison} we analyze the behavior of the index family with two global samples of high resolution inequality data for income- and consumption-based inequality measurements, provided by the World Bank. We also compute the Spearman and Kendall rank correlation matrices of the orderings under multiple different indices. Technical derivations can be found in Appendix \ref{app:technicaldetails} and comparisons with Atkinson- and Amato--Kakwani-indices from Appendix \ref{app:comparisonwithatkinsonandamatokakwaniindices}. Finally, we conclude in Section \ref{sec:conclusions}.

\section{Curvature-based inequality}
\label{sec:curvaturebasedinequality}

For this work, we understand income in an inclusive sense, including various types of monetary resources, the distribution of which can be measured. Later we focus in particular on income and consumption, available from the World Bank Poverty and Inequality Platform \cite{WorldBankPIP}.

\subsection{Lorenz curve and curvature}
\label{subsec:lorenzcurveandcurvature}

The Lorenz curve $\L(x) : [0,1] \to [0,1]$ is defined as \cite{Gastwirth1972, cowell2011measuring}
\begin{equation}
    \L(x) = \frac{1}{\mu} \int_0^x \dd t \ \Q(t) \ ,
\end{equation}
where $\mu$ is the average income and $\Q(t) = \text{inf} \ \qty{x | F(x) \geq t}$ is the quantile function, $F(t)$ is the cumulative income distribution function and $f(t)$ its density function. The Lorenz curve gives for any population fraction $x$ the aggregate amount of income held by such lowest income fraction, and the line of perfect equality is $\L(x) = x$. 

For the purposes of this work, we work only with the classically-defined Lorenz curve, assuming that it has no negative values. We have assumed that the distribution of income has a finite, non-zero mean, as it has in any real society if the income measurement itself cannot be negative or divergent. We do not consider generalizations of the Lorenz curve further in this work.

As usual, we assume the Lorenz curve to be twice-differentiable, and more specifically, we require $\L(0) = 0, \ \L(1) = 1, \ \L \in C^a(0,1)$ with $a \geq 2$. By construction, $\L$ is monotonically increasing and convex. 

As is intuitively clear, the deviation of the Lorenz curve from the line of perfect equality is somehow connected to inequality, and it is up to the index to decide precisely how. Here, we propose measuring inequality based on the curvature of the Lorenz curve. The curvature of any scalar function is defined as the amount the tangent vectors turn as a function of the curve length; such a notion is fundamentally local. This geometric curvature is given by
\begin{equation}
    \kappa(x) = \left| \frac{\dd \phi(x)}{\dd s(x)} \right| \ ,
\end{equation}
where $\phi$ is the rotation angle and $s$ is the curve length. Using
\begin{equation}
    \dd s = \sqrt{1 + (\L'(x))^2} \ \dd x \ , \qquad \dd \phi = \frac{\L''(x)}{1 + (\L'(x))^2} \ \dd x  \ ,
\end{equation}
we have
\begin{equation}
    \kappa(\L(x)) = \frac{\L''(x)}{\qty(1 + (\L'(x))^2)^{3/2}} \ ,
\end{equation}
where absolute values were removed by the convexity of $\L$. 

Note a slight mathematical caveat: assuming a finite population, the Lorenz curve is a piecewise function, and local curvature is ill-defined. Hence, we work with continuous approximations of the Lorenz curve. Finding a fit to data which respects the assumption of convexity is outlined in Appendix \ref{app:technicaldetails}.

\subsection{Inequality from curvature}
\label{subsec:inequalityfromcurvature}

Most standard inequality indices compact inequality into a single scalar value. With a local notion such as $\kappa(\L(x))$, one could imagine a variety of indices where the natural aggregate index is obtained by integrating over the domain. A general form is obtained by considering the full family of $L^p$-norms on this domain. In fact, much like the Atkinson's index (implicitly) using the family of $L^p$-norms with society's inequality-aversion, we define the total curvature $p$-index as\footnote{We could define the index as well by integrating against $\dd s$, the arc length. This would result in an index similar to the one constructed here, and it would also have a closed form at $p = 1$. However, integrating against $\dd x$ weighs each population fraction equally, which seems to be the economically natural measure.}
\begin{equation}
    \I_p = \qty(\int_0^1 \dd x (\kappa(\L(x)))^p)^{1/p} = ||\kappa(\L)||_p \ , \qquad 1 \leq p \leq \infty \ .
\end{equation}
Here, $p = \infty$ corresponds to the supremum norm, $||\kappa(\L)||_\infty = \text{sup}_{x \in [0,1]} \ \kappa(\L(x))$. 

We define society's ``stratification-aversion" parameter to be $\alpha = p - 1$.  As $\alpha \to \infty$, society becomes hypersensitive to stratification; because of the supremum norm, the index measures the maximal value of curvature, which is connected to the maximal differences between income classes. Intuitively, a more stratified society is a more unequal one, and as $\alpha$ is increased, the index punishes income differences increasingly non-linearly.

Note that what we call stratification is a close relative of the concepts of polarization and bipolarization \cite{IncomeDistributionHandbook2014, Esteban1994, Esteban2004, Wolfson1994, Wolfson2010}. Polarization has been axiomatized as well through a so-called identification--alienation framework, considering the clustering of people. The works of Esteban et al. construct a one-parameter family of polarization indices in which a parameter (by chance also denoted $\alpha$) controls for the weight of within-group identification. In a parallel fashion, bipolarization measures primarily the hollowing of the middle class, and has been axiomatized as well. The word stratification has also been used in works related to group overlap \cite{Yitzhaki1991}: there, stratification measures the extent to which groups occupy distinct, non-overlapping strata of an overall distribution. 

Our use of the word assumes no subgroup structure, and stratification is read off from geometry: we use it to denote the existence of sharp changes in distributions, creating differences between population fractions and possibly showing the existence of class boundaries. The curvature-based family can be seen to be reminiscent of polarization indices, but is mostly distinct from them. In particular, studies of polarization often emphasize the difference between (indices of) polarization and inequality, and one of the primary differences is that Pigou--Dalton transfers do not necessarily reduce polarization. As will be seen, with any non-zero $\alpha$, the Pigou--Dalton principle is not guaranteed for the curvature-based index either. 

We have decided to use the word stratification, since we do not attempt to construct an index that would fulfill the polarization index axioms. Also, polarization indices often detect clusters of people concentrated apart, while increasing $\alpha$ rather detects the boundaries between such clusters; the curvature as a local notion can in principle be used to study \textit{where} the divide(s) between classes reside(s). For example, if curvature has multiple local maxima, we could consider society to be stratified into multiple classes, and one should consider the structural factors playing into such formation. Later while studying correlations, we also compare the curvature-based index with a bipolarization index due to Wolfson \cite{Wolfson1994}. 

\begin{figure}[htb]
    \centering
    \includegraphics[width=\linewidth]{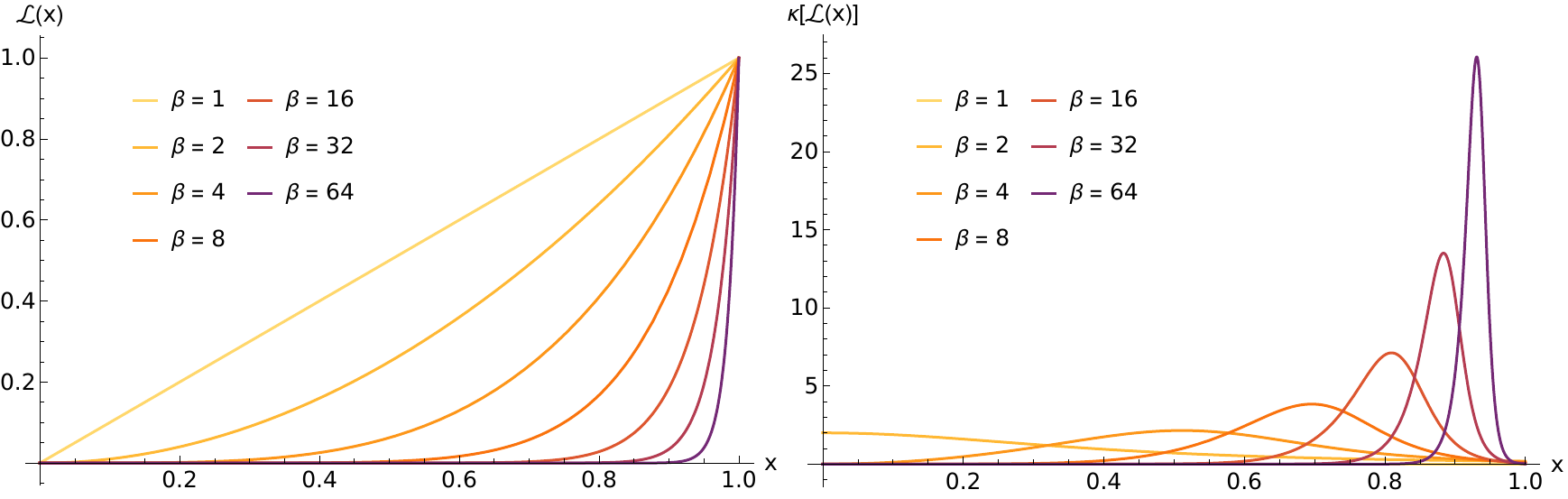}
    \caption{A family of Lorenz curves $x^\beta$ and their curvatures.}
    \label{fig:exampleprofiles}
\end{figure}

\begin{figure}[htb]
    \centering
    \includegraphics[width=0.6\linewidth]{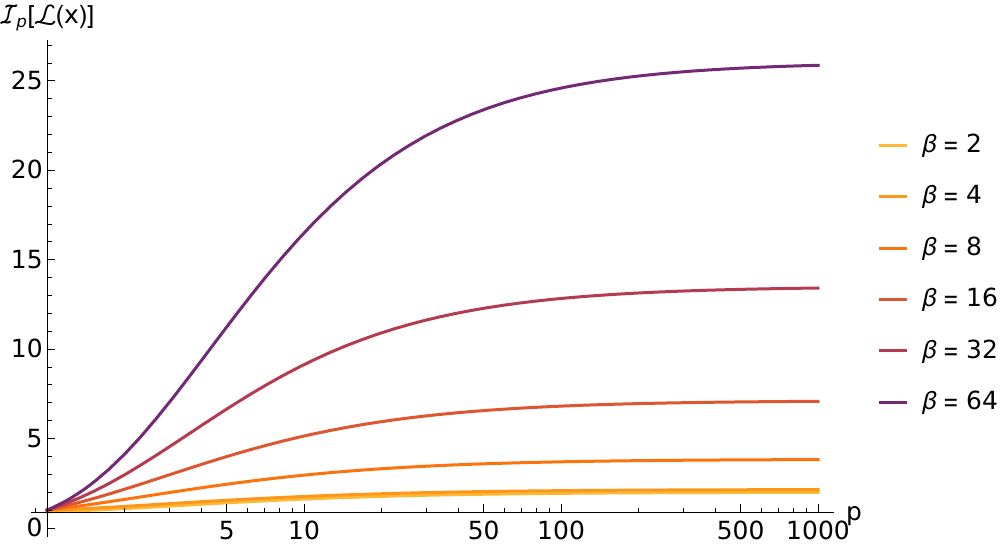}
    \caption{Preliminary index examples as $p$ is varied.}
    \label{fig:exampleindices}
\end{figure}

Putting it all together,
\begin{definition}
    Let $\L :[0,1] \to [0,1]$ be the Lorenz curve, $\L \in C^a(0,1)$ with $a \geq 2$. Its curvature is
    \begin{equation}
        \kappa(\L(x)) = \frac{\L''(x)}{(1 + (\L'(x))^2)^{3/2}} \ .
    \end{equation}
    Define a continuum of inequality and social stratification indices based on this curvature and the family of $L^p$-norms as
    \begin{equation}
        \I_p = \qty(\int_0^1 \dd x \  (\kappa(\L(x)))^p)^{\frac{1}{p}} = ||\kappa(\L(x))||_p \ , \qquad \ 1 \leq p \leq \infty \ .
    \end{equation}
    Choosing $p$ corresponds to choosing society's aversion to stratification, which is measured by the size of local curvature. We define the ``stratification-aversion parameter" as $\alpha = p - 1$.
\end{definition}
Simple examples of curvature for polynomial Lorenz curves of the form $x^\beta$ and the related index values across different values of $p$ are shown in Figures \ref{fig:exampleprofiles}, \ref{fig:exampleindices}.  

As mentioned, the definition of $\I_p$ by curvature is somewhat reminiscent of the definition of the Amato--Kakwani index, defined by the length of the Lorenz curve. Likewise, naming $\alpha$ as stratification-aversion is in accordance with the classical inequality-aversion parameter of the Atkinson index family. Another existing one-parameter index family close to our construction is the single-parameter extended Gini index of Donaldson and Weymark \cite{Donaldson1980} and of Yitzhaki \cite{Yitzhaki1983}, in which an inequality-aversion parameter reweights the gap $x - \L(x)$. See Appendix \ref{app:comparisonwithatkinsonandamatokakwaniindices} for a further comparison between the Atkinson, Amato--Kakwani and the curvature index.


The case of $p = 1$ is special, since we can perform the index integral analytically. Setting $u = \L'(x)$, we find
\begin{equation}
    \I_1 = \int_{\L'(0)}^{\L'(1)}  \frac{1}{(1 + u^2)^{3/2}} \ \dd u= \frac{\L'(1)}{(1 + (\L'(1))^2)^{1/2}} - \frac{\L'(0)}{(1 + (\L'(0))^2)^{1/2}} \ .
\end{equation}
Therefore, with $p = 1$, the value of the index is given directly by the derivatives of the Lorenz curve at the ends of the intervals. Since the Lorenz curve itself is defined as an integral over the quantile function, we have $\L'(1) = y_\text{max}/\mu  \ , \L'(0) = y_\text{min}/\mu$, where $y(x)$ is the income at population fraction $x$. Thus, $\I_1$, originating from a measurement of total curvature, collapses ordinally to a transformed relative range, with a specific functional form. 

As is clear from the closed form, the index is sensitive to interior curvature only as $\alpha > 0$; what $\I_1$ retains from the curvature construction is its specific functional form and extremal tail sensitivity. Below, we show by studying rank correlations that the index is empirically non-redundant with various standard indices, including a standard quantile ratio.   

We may operationalize $\I_1$ in a simple manner by allowing for a tolerance near the ends of the intervals when computing the derivative. For example, a possible sufficient approximation is evaluating the derivatives from $x = 0$ to the 1$^\text{st}$ percentile and from the 99$^\text{th}$ percentile to $x = 1$. That is, one may operationally use
\begin{equation}
    \I_1^\delta = \frac{\L'(1 - \delta)}{\sqrt{1 + (\L'(1-\delta))^2}} - \frac{\L'(\delta)}{\sqrt{1 + (\L'(\delta))^2}} \ ,
\end{equation}
with $\delta$ small enough to remain near the tails, yet large enough to be stably estimable from percentile data. This form is more robust to income distribution tail noise, and easily computable from percentile share data without any fitting. For theoretical income distribution models with unbounded support and zero minimum income, the continuous version of the index saturates to one. In such cases only the operationalized $\I^\delta_1$ is informative. Naturally, this poses no problems with real data.

Additionally, studying the World Bank inequality dataset for 172 countries, we have found, using the convex fits described in Appendix \ref{subapp:maintainingconvexityfordifferentiablelorenzcurves}, that 23 of the 172 countries have a curvature maximum inside the interval instead of exactly at endpoints. This motivates one to consider what would be a suitable non-zero $\alpha$ to capture such a structural feature.

The promise of $\I_p$ is to unify the measurement of inequality and social stratification from the Lorenz curve by defining the index as a continuum parametrized by $\alpha$. As emphasized in polarization literature, there is a distinction between inequality and polarization. If one considers the $\I_p$ family at the two extremes to measure either only inequality or only stratification, we have constructed a continuum between the two quantities of interest. However, we have not yet discussed any of the properties an axiomatically pure inequality index must fulfill, which we shall do now.

\section{Axiomatic properties}
\label{sec:axiomaticproperties}

Let us study the fulfillment of axiomatic properties within the $\I_p$-family. As will be seen, $p = 1$ (or $\alpha = 0$) is special in more ways than just having a closed form; we divide the following exposition into the cases $p = 1$ and $p > 1$. As a preceding summary, we will find that all axiomatic properties hold for the curvature index $\I_p$ in general if and only if $p = 1$ (Pigou--Dalton transfer principle holding only weakly). If $p > 1$, that is, if society has any non-zero amount of aversion to stratification, normalizability, Pigou--Dalton transfer principle and Lorenz dominance do not necessarily hold. Each mode of failure is given a more technical exposition either in the main text or in Appendix \ref{app:technicaldetails}.

\subsection{Anonymity, population invariance and scale invariance}
\label{subsec:anonymitypopulationinvarianceandscaleinvariance}

Since the $\I_p$-family is based on the Lorenz curve which by construction is anonymous and invariant under the scaling of population or income, also the indices inherit these properties.

\subsection{Normalizability}
\label{subsec:normalizability}

An often-stated requirement is that an inequality index must be normalized between 0 and 1, but what is really wanted is normalizability. For a purely ordinal index, even normalizability is not necessary. Of course, performing normalization can allow a uniform view of different indices normalized the same way (canonically between 0 and 1), setting up a standardized frame of reference.

By the Hölder inequality, $\I_p$ is a non-decreasing function of $p$; in particular, with $q \geq p$
\begin{equation}
    ||\kappa||_p^p \leq \qty(\int_0^1 ((\kappa(x))^p)^{q/p} \dd x)^{p/q} \cdot \qty(\int_0^1 1^\beta \dd x )^{1/\beta} =  ||\kappa||_q^p \ ,
\end{equation}
with $\beta = \frac{q}{q-p}$. Taking the $p^\text{th}$ root on both sides shows that for any $q \geq p$, $||\kappa||_p \leq ||\kappa||_q$. Thus, unless $\kappa(\L(x))$ is a constant, increasing stratification-aversion always increases measured inequality.

An interested reader may find the mathematically-inclined normalizability discussion in Appendix \ref{subapp:normalizability}. The punchline is: with $p = 1$, normalizability is satisfied -- in fact, the index is already normalized. In the extreme case where one person owns everything, $\L'(1) \to \infty, \ \L'(0) = 0$, and with $p = 1$,
\begin{equation}
    \I_1(``\delta(1)") = \lim_{x \to \infty} \frac{x}{\sqrt{1 + x^2}} - 0 = 1 \ .
\end{equation}
For higher $p$, the situation is subtle. Measures like Gini automatically cap inequality at 1 for a society where one person owns everything, while curvature-based measures with \textit{any} amount of stratification-aversion recognize that the inequality\footnote{Since at $\alpha > 0$ the index is not an axiomatically pure inequality measure, in that regime $\I_p$ is better read as a combined measure than of inequality proper.} of such a society can be arbitrarily large.

Since by definition $\L(x)$ is at least twice-differentiable, a single country always has a finite value of $\I_p$ for any $p$. However, finiteness does not imply normalizability; as a functional, $\I_p$ is fundamentally unbounded. For any real finite sample of countries, we may still normalize with respect to the largest value of inequality found in such sample. An ordinal comparison is always readily available for any $p$ without transformations.

We conclude that normalization and more importantly, normalizability, holds with $p = 1$, but does not hold for any $p > 1$ in the strict sense. Normalization can still be performed for a finite sample of countries if needed.

\subsection{Pigou--Dalton transfer principle}
\label{subsec:pigoudaltontransferprinciple}

The Pigou--Dalton transfer principle asserts that any rank-preserving transfer from a richer to a poorer individual should decrease inequality \cite{Dalton1920}. Strictly speaking, the transfer principle at its bare formulation is meaningless for curvature-based indices, since curvature is defined only in the continuous approximation. Hence, we study the principle by studying the effect of continuous deformations of the Lorenz curve and its effect on inequality. The full example computation can be found in the technical appendix \ref{subapp:pigoudaltontransferprinciplefromacontinuousvariation}, while here we only state the main results.

A continuous progressive transfer can be modeled as a small positive perturbation $\phi(x)$ added to the Lorenz curve. For example, such a perturbation necessarily reduces the area-based Gini coefficient, but its effect on curvature is not straightforward.

The case of $p = 1$ is again special: One may argue that a transfer from a richer individual to a poorer either reduces $\L'(1)$ or keeps it as is, and either increases $\L'(0)$ or keeps it as is. Hence, the index family satisfies the Pigou--Dalton principle, but only in a weak sense: for example, a transfer inside the middle class may not move either derivative, keeping measured inequality the same.

The argument one may give with general $p$ is quite limited. Since the condition derived in Appendix \ref{subapp:pigoudaltontransferprinciplefromacontinuousvariation} is only sufficient but not strictly necessary, various other types of transformations can also reduce inequality with less stringent forms. Intuitively, if $p \neq 1$, a transfer may go from a richer to a poorer individual, but at the same time it can strengthen the stratification by making some other class-barrier higher. This change is captured and punished by a non-zero stratification-aversion parameter, and as a result, such transfer does not necessarily reduce measured inequality (analogous to polarization). Of course, welfare programs are often also aimed at lessening stratification, and hence they can be considered likely to reduce inequality as measured by $\I_p$ even as $p > 1$.

\subsection{Lorenz dominance}
\label{subsec:lorenzdominance}

Finally, Lorenz dominance asserts that any Lorenz curve which is pointwise larger for all $x \in [0,1]$ than another Lorenz curve results in a less unequal society. For traditional indices like the area-based Gini this property is again immediate.

With $p = 1$ Lorenz dominance is manifest: If a Lorenz curve lies above another Lorenz curve, it must have both smaller slope at $x = 1$ and larger slope at $x = 0$, making its curvature inequality smaller: let $\L_1 \geq \L_2 \ \forall \ x \in [0,1]$ and consider the limit $x \to 1$. By dominance, 
\begin{equation}
    \L_1(x) - \L_2(x) \geq 0 \ ,
\end{equation}
so as $x \to 1$,
\begin{equation}
    \frac{\L_1(1) - \L_1(x)}{1-x} - \frac{\L_2(1) - \L_2(x)}{1-x} \leq 0 \ ,
\end{equation}
and taking the limit $x \to 1$
\begin{equation}
    \L_1'(1^-) \leq \L_2'(1^-) \ .
\end{equation}
The limit $x \to 0^+$ is analogous by symmetry.

\begin{figure}
    \centering
    \includegraphics[width=\linewidth]{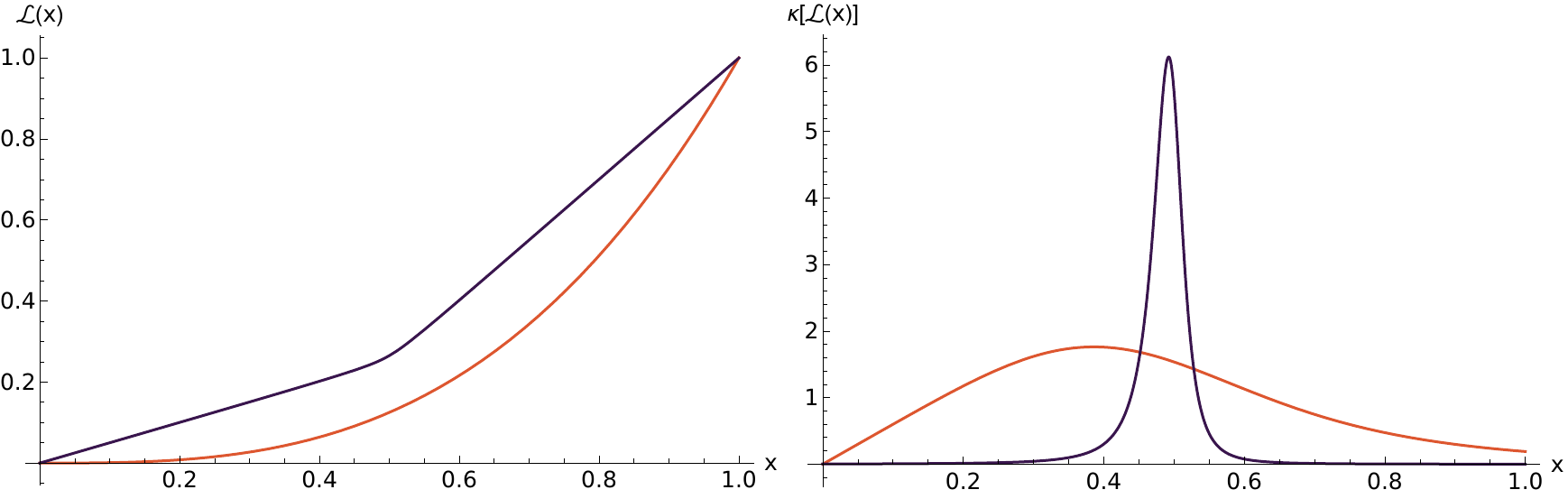}
    \caption{Two Lorenz curves: $\L_2$ Lorenz-dominates $\L_1$, yet attains a much higher maximum curvature.}
    \label{fig:dominanceexample}
\end{figure}

As an example for the failure of this condition for general $p$, consider two societies, a smoothly varying yet unequal society described by the Lorenz curve $\L_1(x) = x^3$, and an \textit{a priori} more equal society with a sharp class transition, Lorenz-dominating the first. We construct this curve with hyperbolic smoothing,
\begin{equation}
    \L_2(x) = x + 0.5 \cdot \qty(\sqrt{(x-0.5)^2 + \epsilon} - \sqrt{0.25 + \epsilon}) \ .
\end{equation}
For demonstration, set $\epsilon = 10^{-3}$. $\L_{1,2}$ and their curvatures are shown in Figure \ref{fig:dominanceexample}. The indices $\I_p$ for these curves are equal at $p \approx 1.87$.
In general, for any Lorenz curves $\L_1, \ \L_2$ with $\I_1(\L_1) < \I_1(\L_2)$, but where $\L_1$ has a larger global curvature maximum than $\L_2$, there exists a $p$ after which the first curve is measured to be more unequal than the second; the functional evaluating inequality is continuous, and at $p \to \infty$ only maximal curvature matters, so by the intermediate value theorem there exists a crossover of the two measures. 

\begin{figure}[ht]
    \centering
    \includegraphics[width=0.6\linewidth]{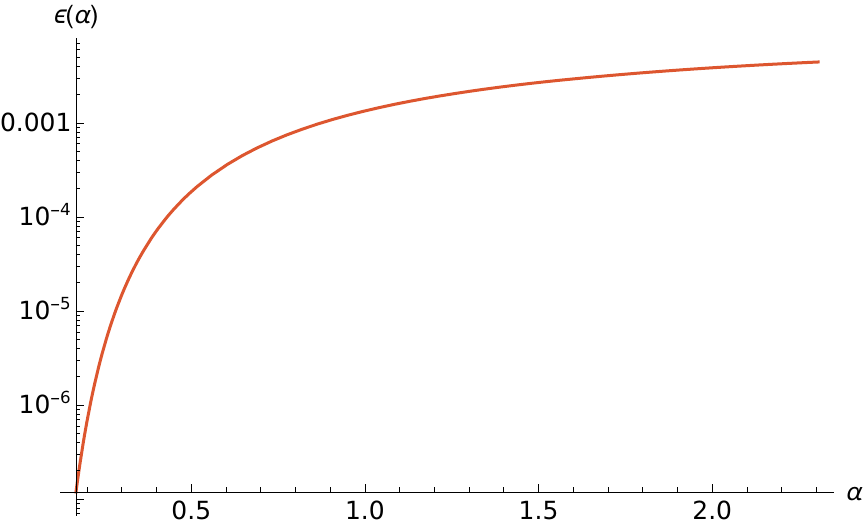}
    \caption{A sharpness of the transition $\epsilon$ which results in the same level of inequality with a given stratification-aversion. As $\alpha \to 0$, the transition can be arbitrarily sharp --- $\I_1$ respects Lorenz dominance.}
    \label{fig:pepsplot}
\end{figure}

As a final visualization of this, we can find $\epsilon(\alpha)$ for the two Lorenz curves above, which shows for a given level of stratification-aversion, at which transition sharpness $\epsilon$ the two societies are measured as equally unequal. We numerically solve this function and show it in Figure \ref{fig:pepsplot}.

\subsection{Axiomatic summary}
\label{subsec:axiomaticsummary}

Summarizing, we have found that all axiomatic properties hold for the curvature index $\I_p$ in general if and only if $p = 1$, where Pigou--Dalton holds only weakly. If $p > 1$, that is, if society is even a bit averse to stratification, normalizability, Pigou--Dalton transfer principle and Lorenz dominance do not hold in general. 

Similarly to Atkinson's index considering how well the Rawlsian maximin principle is followed, the $\I_p$-family considers how much stratification or more concretely, boundaries between income classes can be accepted. For example, a society can have a Gini index arbitrarily close to 0, but still have a point of large curvature, showing a sharp (albeit perhaps meaningless) class-divide. It is up to another study to find how much stratification-aversion should be considered reasonable. After all, income differences and indirectly stratification-aversion are a recurring question in everyday politics, and could perhaps be inferred from answers to surveys or other large datasets.

The possibility of failure of Lorenz dominance and Pigou--Dalton transfers for $\alpha > 0$ should be considered in light of the polarization literature \cite{Esteban1994, Esteban2004}. For example, equalizing within a group is a sequence of Pigou--Dalton transfers, reducing inequality for any Lorenz-consistent index, while strengthening group identification and thus increasing polarization. With the stratification-aversion parameter, a similar story applies for Lorenz dominance. Summarized, any measure sensitive to distributional clustering or the walls resulting from such clustering must give up Lorenz consistency by design. The curvature-family $\I_p$ makes this trade-off explicit and continuous: any $\alpha > 0$ gains sensitivity to stratification at the cost of perfect Lorenz consistency in analogy with how much society is averse to polarization.

\section{Global prospects and comparison}
\label{sec:globalprospectsandcomparison}

Having established the one-parameter family of Lorenz curvature-based, combined inequality and stratification indices, we move on to an empirical examination. In this section we compute different-$p$ versions of the proposed indices for a global selection of countries, showcasing the effect of changing $\alpha$, and compare the values of these indices to other standard measures of inequality. We also compute the Spearman and Kendall rank correlation matrices between $\I_1$ and other common indices for the full selection of income- and consumption-based inequality data.

\subsection{Global comparison}

To study the indices reliably, we need data to be granular enough for a faithful reconstruction of the underlying Lorenz curve. We have used the World Bank's PIP \cite{WorldBankPIP, WorldBankPIPData} datasets for percentile shares, from which Lorenz curves have been reconstructed. We have collected the percentile data, choosing the latest yearly dataset available for each nation. 

The data itself has been collected based on either income or consumption. Especially in agrarian and less developed economies, income may be quite volatile and hard to measure, so consumption acts as a proxy for the actual standard of living \cite{OxfordHandbook}. We split the analysis into two sets based on the distinction between income and consumption.

For the first visualizations, we have collected a specific sample of nations with a wide global coverage. The differing qualitative features range from geographical location, the usual perceived level of inequality and internal stratification, speed of growth and development, historical societal transitions, different types of economies, and simply the size of the nation, to name a few. Based on these, we have chosen the following 12 nations for each group:
\begin{itemize}
    \item Inequality measured by income: Australia, Brazil, Chile, Colombia, France, Germany, Mexico, Poland, Sweden, Turkey, United Kingdom and United States.
    \item Inequality measured by consumption: China, Egypt, Ethiopia, India, Indonesia, Côte d'Ivoire, Kenya, Morocco, Nigeria, South Africa, Thailand and Vietnam. 
\end{itemize}

We outline the numerical method used in the computation in the technical appendix. Due to data sparsity, continuous methods are numerically challenging to utilize for $\alpha \neq 0$, and the visualization should be considered representative mostly qualitatively; see Appendix \ref{subapp:maintainingconvexityfordifferentiablelorenzcurves} for details.

\begin{figure}[htb]
    \centering
    \includegraphics[width=0.8\linewidth]{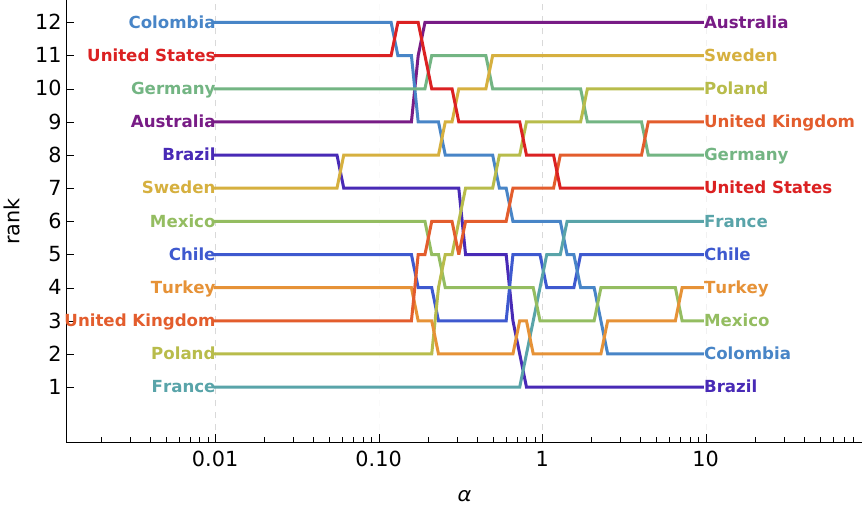}
    \caption{Comparison of ranking order as $\alpha$ is varied between $[0.01, 10]$ for income-based countries in the sample. $y$-axis corresponds to ranking, so that the most equal country is at the bottom.}
    \label{fig:incomealpha}
\end{figure}

\begin{figure}[htb]
    \centering
    \includegraphics[width=0.8\linewidth]{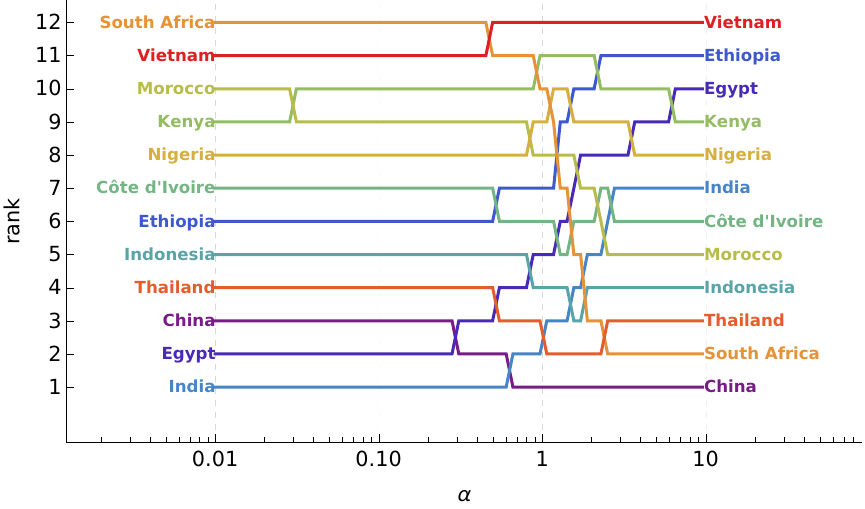}
    \caption{Comparison of ranking order as $\alpha$ is varied between $[0.01, 10]$ for consumption-based countries in the sample. $y$-axis corresponds to ranking, so that the most equal country is at the bottom.}
    \label{fig:consumptionalpha}
\end{figure}

The resulting country orderings are shown in Figures \ref{fig:incomealpha}, \ref{fig:consumptionalpha}. Multiple swaps between countries take place as $\alpha$ varies between $0.01$ and $10$. We have studied the transition further in $\alpha$ in both directions, and the rankings are stable outside this window. 

A few remarks are in order: Compared to usual indices, Sweden is ranked considerably low among income-based nations, while China is ranked considerably high among consumption-based nations. As noted, $\I_1$ can be cast as a function of tail differences, and both low bottom-incomes and high top-incomes are punished by curvature-based indices in ranking. This is one of the features many other indices wash away when averaging over the derivatives.

Second, looking at the transitions of countries as $\alpha$ is varied, we see that income-based countries start changing positions on average with lower $\alpha$ than consumption-based countries. This might be a mirage due to low sample size, but an effect worth considering further with larger data and sharper estimates. Should it be a true feature, it could signal, for example, that consumption-based measures measure less stratification fundamentally, or that the consumption-based countries are on average less stratified, as measured by curvature.

\subsection{Index comparison}
\label{subsec:indexcomparison}

We also compare $\I_1$ ordinally with various standard inequality indices in their continuous formulation for the above set of countries. For comparison, we have chosen the Gini, Atkinson ($\epsilon = 0.5$), Amato--Kakwani, and Theil-1 indices, as well as the standard quantile ratio $P90/P10$ and the polarization index due to Wolfson. Their Lorenz curve-based formulations are \cite{Gini1912book, Atkinson1970, Amato1968, Kakwani1980, Theil1967, Wolfson1994, cowell2011measuring}
\begin{equation}
\label{eq:otherindices}
    \begin{split}
        G &= 1 - 2 \int_0^1 \L(x) \ \dd x \\
        A_{0.5} &= 1 - \qty(\int_0^\infty \qty(\frac{y}{\mu})^{\frac{1}{2}} f(y) \ \dd y)^2 = 1 - \qty(\int_0^1 \sqrt{\L'(x)} \ \dd x)^2 \\
        K &= \frac{1}{2  - \sqrt{2}}\qty(\int_0^1 \sqrt{1 + (\L'(x))^2} \ \dd x -\sqrt{2}) \\
        T_1 &= \int_0^\infty \frac{y}{\mu} \ln \qty(\frac{y}{\mu}) f(y) \ \dd y = \int_0^1 \L'(x) \ln \qty(\L'(x)) \ \dd x  \\
        P90/P10 &= \frac{\L'(0.9)}{\L'(0.1)}  \\
        W &= 2\frac{1 - 2 \L(0.5)  -  G}{\L'(0.5)} \ .
    \end{split}
\end{equation}
We compute each of these indices for the two samples of countries discussed above, and the results are shown in Tables \ref{tab:incomecomparison} and \ref{tab:consumptioncomparison}. Without additional normalization absolute values are not comparable, so we compare ordinally. 
\begin{table}[htb]
    \centering
    \begin{tabular}{c|c|c|c|c|c|c|c}
    Country & $\I_1$ rank & $G$ rank & $A_{0.5}$ rank & K rank & $T_1$ rank & $\frac{P90}{P10}$ rank & Wolfson rank\\
    \hline \hline 
     Australia & 9 & 6 & 5 & 5 & 5 & 6 & 6 \\
     Brazil & 8 & 11 & 11 & 11 & 11 & 11 & 11 \\
     Chile & 5 & 9 & 9 & 9 & 9 & 7 & 7 \\
     Colombia & 12 & 12 & 12 & 12 & 12 & 12 & 12 \\
     France & 1 & 3 & 3 & 3 & 3 & 3 & 3 \\
     Germany & 10 & 5 & 6 & 6 & 6 & 5 & 4 \\
     Mexico & 6 & 8 & 8 & 8 & 8 & 8 & 8 \\
     Poland & 2 & 1 & 1 & 1 & 1 & 1 & 1 \\
     Sweden & 7 & 2 & 2 & 2 & 2 & 2 & 2 \\
     Turkey & 4 & 10 & 10 & 10 & 10 & 9 & 9 \\
     United Kingdom & 3 & 4 & 4 & 4 & 4 & 4 & 5 \\
     United States & 11 & 7 & 7 & 7 & 7 & 10 & 10\\
\end{tabular}
    \caption{Rankings of countries with an income-based Lorenz curve under various indices.}
    \label{tab:incomecomparison}
\end{table}

\begin{table}[htb]
    \centering
    \begin{tabular}{c|c|c|c|c|c|c|c}
    Country & $\I_1$ rank & $G$ rank & $A_{0.5}$ rank & K rank & $T_1$ rank & $\frac{P90}{P10}$ rank & Wolfson rank\\
    \hline \hline 
     China & 3 & 8 & 8 & 8 & 8 & 8 & 9 \\
     Egypt & 2 & 2 & 2 & 2 & 2 & 2 & 2 \\
     Ethiopia & 6 & 3 & 3 & 3 & 3 & 3 & 3 \\
     India & 1 & 1 & 1 & 1 & 1 & 1 & 1 \\
     Indonesia & 5 & 6 & 6 & 6 & 7 & 4 & 4  \\
     Côte d'Ivoire & 7 & 7 & 7 & 7 & 6 & 7 & 8 \\
     Kenya & 9 & 10 & 10 & 10 & 10 & 10 & 11 \\
     Morocco & 10 & 11 & 11 & 11 & 11 & 11 & 10 \\
     Nigeria & 8 & 5 & 5 & 5 & 5 & 6 & 5 \\
     South Africa & 12 & 12 & 12 & 12 & 12 & 12 & 12 \\
     Thailand & 4 & 4 & 4 & 4 & 4 & 5 & 6 \\
     Vietnam & 11 & 9 & 9 & 9 & 9 & 9 & 7 \\
\end{tabular}
    \caption{Rankings of countries with a consumption-based Lorenz curve under various indices.}
    \label{tab:consumptioncomparison}
\end{table}

For this sample, the curvature index agrees to an extent with usually-employed indices, which are largely unanimous in their ranking. The rankings given by the curvature index are quite comparable to the other indices, except for a few surprising rankings; in particular, the low rank of Sweden among income-measured countries and the high rank of China among consumption measured countries, as already mentioned. Even so, as was shown, $\I_1$ is an axiomatically pure index (modulo Pigou--Dalton being only weakly satisfied).

\begin{figure}[htb]
    \centering
    \includegraphics[width=\linewidth]{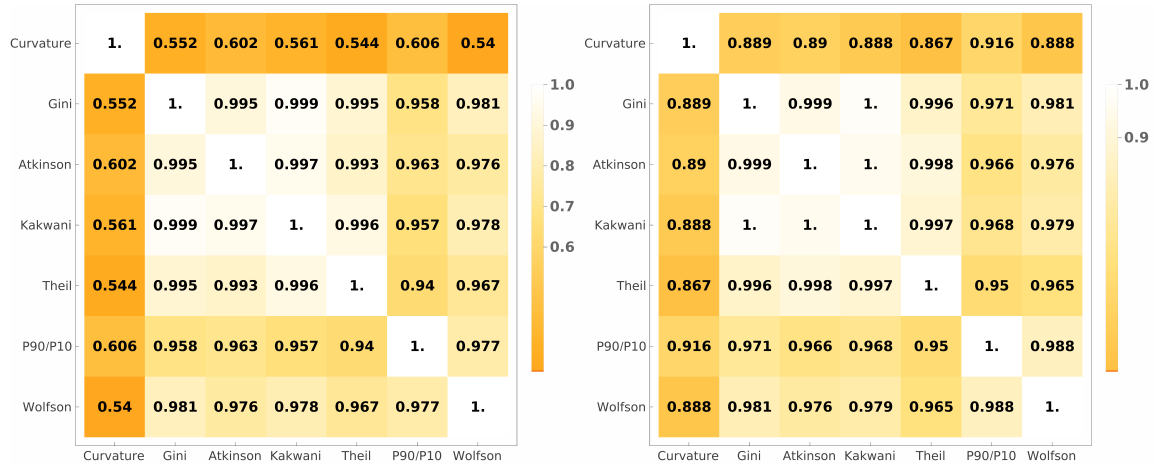}
    \caption{Spearman rank correlations for the full set of income-based (left) and consumption-based  (right) countries in the PIP dataset for the seven different indices considered.}
    \label{fig:correlationmatricesSpearman}
\end{figure}

\begin{figure}[htb]
    \centering
    \includegraphics[width=\linewidth]{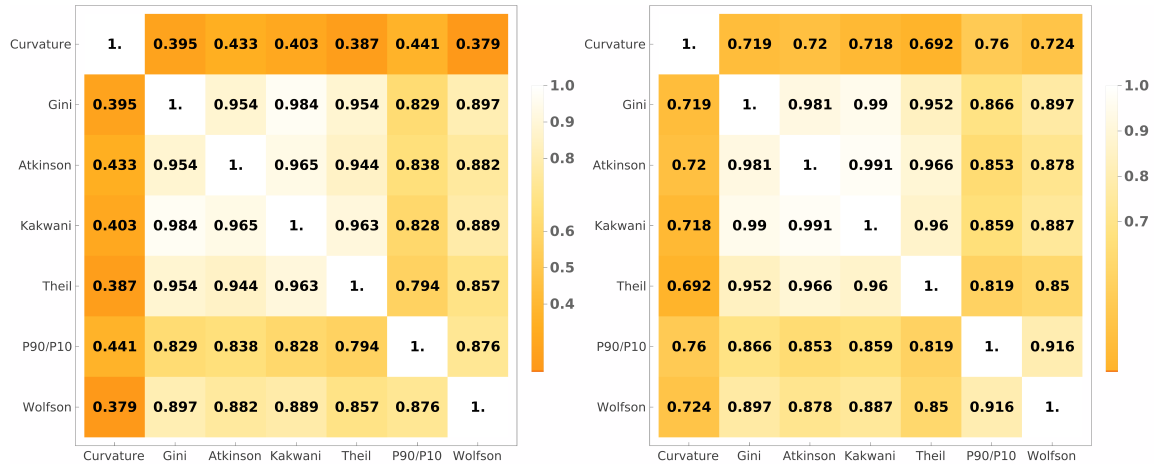}
    \caption{Kendall rank correlations for the full set of income-based (left) and consumption-based (right) countries in the PIP dataset for the seven different indices considered.}
    \label{fig:correlationmatricesKendall}
\end{figure}

As a final check, we have extracted the latest inequality data for all countries in the PIP dataset, again grouping by income and consumption. The full dataset includes 72 income-based countries and 100 consumption-based countries. For both sets, we construct an ordering based on each of the seven indices. From these ordering vectors, we compute the Spearman and Kendall rank correlations, and gather them into correlation matrices. These matrices are shown in Figures \ref{fig:correlationmatricesSpearman} and \ref{fig:correlationmatricesKendall}, respectively.

As already hinted by the small sample above, classical indices like Gini, Theil, Atkinson and Amato--Kakwani are extremely correlated as measured by either Spearman or Kendall. The Spearman correlations of the quantile ratio and the Wolfson index with each other and the above four cross-correlated indices are also very high, while Kendall correlations are somewhat lower, yet still large. 

Across the board, the curvature-based index shows a meaningful deviation with respect to all other indices considered. As described in Appendix \ref{subapp:maintainingconvexityfordifferentiablelorenzcurves}, we have tested the robustness of the Kendall correlation differences with order-15 and order-25 polynomial fits, and found them to be approximately on par with the order-20 fits. Hence, the ranking difference compared to other indices is not due to numerical noise. $\I_1$ correlates somewhat more strongly with the quantile ratio than the other indices. Notably, rank correlations of the curvature index with the other six are much larger for consumption-based countries than income-based countries, as measured by either type of correlation.

\section{Conclusions}
\label{sec:conclusions}

In this work we have constructed a novel family of combined inequality and social stratification indices based on the geometric curvature of the Lorenz curve. The index family was parametrized by the so-called stratification-aversion parameter $\alpha$, which determined how much large curvatures of the Lorenz curve were punished by the index.

We considered the axiomatic purity of the family of indices, also showcasing where these requirements fail. Anonymity, population invariance and scale invariance were immediate, while normalizability, Pigou--Dalton transfers and Lorenz dominance required further study. It was found that with $p = 1$ ($\alpha = 0$), all of the axioms were satisfied (Pigou--Dalton only weakly), while with $p > 1$ we could always construct pathological cases resulting in these conditions failing. These findings were connected to models and indices on polarization. 

Finally, we studied in detail the actual behavior and performance of the index. Using percentile share data from the PIP dataset, we showcased for a sample of income-based and a sample of consumption-based inequality data how varying $\alpha$ results in varying rankings given by the index. After this, using $p = 1$, we computed the ordinal rankings of this sample and compared it to the ordinal rankings given by six different indices: the Gini, Atkinson with $\epsilon = 0.5$, Amato--Kakwani and the Theil-1 index, as well as the $P90/P10$-quantile ratio and the Wolfson index. Finally, we computed the Spearman and Kendall rank correlation matrices for the seven indices from the full set of countries available in PIP data. 

Multiple lines of generalization lie open. As was noted, the existence of the second derivative in the index was fundamental in being able to track stratification for $\alpha > 0$ in a society, and the appearance of this second derivative was very natural, resulting directly from curvature. However, there seems to be no obvious limitation to constructing indices dependent on higher derivatives as well, albeit in that case the interpretation of a parameter like $\alpha$ should also be updated. 

Another straightforward generalization would be to allow computing this index also on Lorenz curves with negative income. This of course first and foremost requires a generalization of the Lorenz curve itself, some of which exist in the literature, along with generalizations of indices \cite{ChauNan1982}. Allowing for negative incomes should result in higher values of curvature, and the indices would require a normalizing transformation, should this line of study be pursued. 

Finally, various dynamical economic models can be studied, each of which generate families of Lorenz curves. The curve families generated for example as solutions to partial differential equations describing income dynamics can be fit to data, such that the source of curvature is directly tied to economics-related parameters from the original equations. We leave these investigations for possible future work.

\section*{Acknowledgements}

The author gratefully acknowledges Aleksi Piispa, not only for their early conceptual ideas but also for their valuable perspectives and intellectually enriching conversations.

\section*{Data and code availability}
The analysis code and processed
per-country Lorenz-curve data are openly available at
\url{https://github.com/ahippelainen/curvature_inequality}. The underlying
percentile shares are from the World Bank Poverty and Inequality
Platform \cite{WorldBankPIP, WorldBankPIPData} (CC-BY 4.0).

\begin{appendix}

\section{Technical details}
\label{app:technicaldetails}

Let us expand on the technical details of the computations which were outside the scope of the main text.

\subsection{Continuous approximations of Lorenz curves}
\label{subapp:maintainingconvexityfordifferentiablelorenzcurves}

Even though data is reported at percentile level, for the purposes of numerical computation it is still quite sparse. Especially for comparisons with respect to $\alpha$, we need a stable-enough approximation of data. We must ensure the standard properties of a Lorenz curve are satisfied: $\L(0) = 0, \ \L(1) = 1$, and that $\L$ stays positive and convex. A direct spline-interpolation fails this due to sparsity.

A particularly useful basis for this fit is given by the Bernstein polynomials, 
\begin{equation}
    B_{k,M}(x) = \binom{M}{k} x^k (1-x)^{M-k} \ ,
\end{equation}
such that
\begin{equation}
     \L(x) = \sum_{k=0}^M c_k B_{k,M}(x) \ .
\end{equation}
An important property of Bernstein polynomials is that we can control both the monotonicity and convexity of the fit \cite{DeVore1993}. In particular, to obtain $\L(0) = 0$ and $\L(1) = 1$, we require $c_0 = 0, \ c_M = 1$. For monotonicity we constrain $c_k \leq c_{k+1}$, and to ensure convexity, constrain $c_{k+2} - 2 c_{k+1} + c_k \geq 0$. The actual fit minimizes squared residuals with respect to data, given the constraints.

For numerical evaluation of indices, we sample the fitted Lorenz curves on a Gauss-Lobatto grid, after which we compute derivatives of the functions with pseudospectral differentiation matrices. This provides stability and efficient evaluation.

For the curvature index $\I_1$ we need only derivatives at the ends of the interval. There are multiple methods of computing these without requiring a full fit to a functional form, but without a full fit we lose control on convexity. The simplest such approximation is a finite difference derivative (FDD) with a given order of approximation. However, we find this to give unsatisfactory results; in particular, some first derivatives at $x = 0$ are slightly negative due to data sparsity.

Another approximation is based on a method originally due to Savitzky and Golay \cite{Savitzky1964}. At both ends of the interval, we fit a low order polynomial to a small set of points, from which we compute the derivatives. Fitting a cubic polynomial to the first and last 13 points of the interval, this method fares better than a direct FDD computation, but a small number of derivatives are still slightly negative at $x = 0$. 

We have computed the ordering of countries with both schemes of low order polynomial fitting and computing the derivatives directly from the convex Bernstein polynomial fit. For the income-based country sample, two pairs of countries switch positions with changing scheme, while all consumption-based countries have exactly the same ranking under both schemes. Hence, we use the convex Bernstein polynomial fits in evaluating the indices. As discussed in the main text, the simplest operational choice would be to compute the derivatives directly from the first and last percentile points. 

For the fits we have used an order-20 Bernstein polynomial across the board, but tested also for robustness in varying the fitting order. For the 12 income- and consumption-based country samples, we have found that changing the order to 15 or 25 sometimes results in ordering swaps, but only minor ones (at most two ranks, most often no swap). Hence, we have decided to use the continuous formulation instead of the operationally easier percentile-formulation to conserve convexity.

\subsection{Normalizability}
\label{subapp:normalizability}

\begin{figure}
    \centering
    \includegraphics[width=\linewidth]{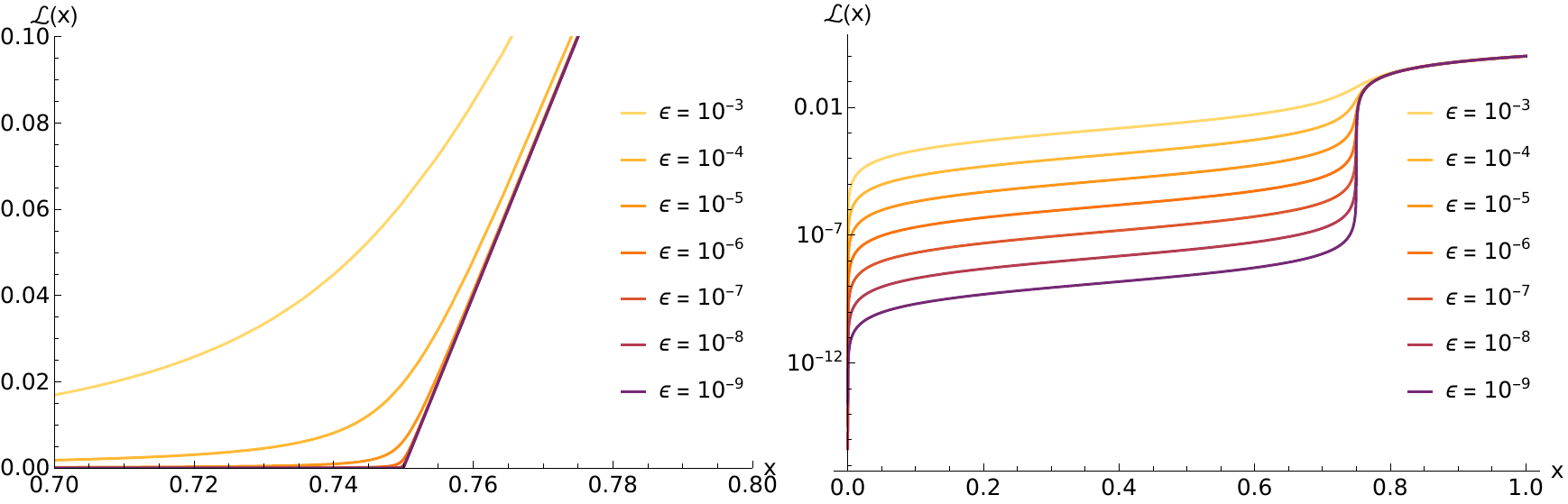}
    \caption{A sequence of kinked Lorenz curves.}
    \label{fig:lorenzkinks}
\end{figure}

\begin{figure}
    \centering
    \includegraphics[width=0.8\linewidth]{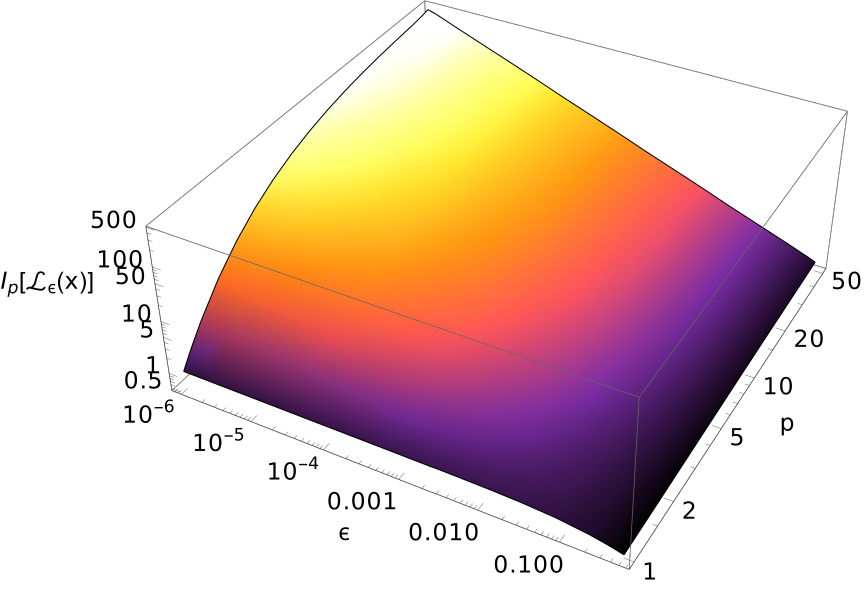}
    \caption{Divergences of $\I_p$ with kinked profiles. Only $p = 1$ stays finite with decreasing $\epsilon$.}
    \label{fig:indexdivergences}
\end{figure}

For completeness, let us quantify how normalizability can break. Consider a sequence of increasingly ``kinked" Lorenz curves,
\begin{equation}
\begin{split}
    F_\epsilon(x) &= x - x_0 + \sqrt{\qty(x - x_0)^2 + \epsilon} \\
    \L_\epsilon(x) &= \frac{F_\epsilon(x) - F_\epsilon(0)}{F_\epsilon(1)-F_\epsilon(0)} = \frac{x + \sqrt{(x-x_0)^2 + \epsilon} - \sqrt{x_0^2 + \epsilon}}{1 + \sqrt{(1-x_0)^2 + \epsilon} - \sqrt{x_0^2 + \epsilon}} \ .
\end{split}
\end{equation} 
An example is depicted with $x_0 = \frac{3}{4}$ in Figure \ref{fig:lorenzkinks}. We also compute the index with various combinations of $p$ and $\epsilon$ and plot these in Figure \ref{fig:indexdivergences}.

Let us derive at which $p$, given such a kink, the curvature index starts to diverge as $\epsilon \to 0$. Assume $0 \neq x_0 \neq 1$, both of which correspond to degenerate cases. \textit{A priori}, it is possible that some of the kink singularities would be integrable, still resulting in a finite index. Since $\lim_{\epsilon \to 0} F_\epsilon(x)$ is finite, the divergences result only from derivatives, and divergences of $\L(x)$ depend directly on derivatives of $F_\epsilon$. Compute these and change variables $x - x_0 = \sqrt{\epsilon} u$,
\begin{equation}
    F_\epsilon'(u) = 1 + \frac{u}{\sqrt{1 + u^2}} \ , \qquad F_\epsilon''(u) = \frac{1}{\epsilon^{1/2}}\frac{1}{(1 + u^2)^{3/2}}\ .
\end{equation}
Thus, the possible divergence comes from the term $\epsilon^{\frac{1-p}{2}}$ in front of the curvature integral, with $\epsilon^{1/2}$ coming from the change of measure and the rest of the integral being finite. The index stays finite as $\epsilon \to 0$ if
\begin{equation}
    \frac{1- p}{2} \geq 0 \iff p \leq 1 \ ,
\end{equation}
but $p \in [1,\infty)$. Therefore, if $p > 1$, a Lorenz curve with such a kink provides a diverging curvature integral. Naturally, the critical $p$ is curve-specific, as is whether it falls within the admissible range $p \in [1, \infty)$. For example, kinks smoothed with a larger power may admit a range of values of $p$ which keep the index generally bounded.

\subsection{Pigou--Dalton transfer principle from a continuous variation}
\label{subapp:pigoudaltontransferprinciplefromacontinuousvariation}

Consider adding a bump function $\phi(x)$ to the Lorenz curve. Let $\phi(x)$ be a compactly supported test function, $\phi(x) \in C^\infty_c(a,b)$ for some interval $[a,b] \subset [0,1]$. In particular, to ensure that the varied curvature has no discontinuities at any point and that Lorenz curve boundaries are respected, impose the boundary conditions
\begin{equation}
\label{eq:phiboundaryconditions}
\phi(a) = \phi'(a) = \phi''(a) = 0 \ , \qquad \phi(b) = \phi'(b) = \phi''(b) = 0 \ .
\end{equation}
We require the transfer to be progressive, that is, for all $x, \ \phi(x) \geq 0$, and strictly positive at some point on the interval.  Let the modified Lorenz curve be
\begin{equation}
    \L_\epsilon(x) = \L(x) + \epsilon \phi(x) \ ,
\end{equation}
where $\epsilon > 0$ and small. The functional variation is
\begin{equation}
    \delta \I_p^p[\kappa[\L_\epsilon]] = \int_0^1 \pdv{\kappa^p}{\kappa} \delta \kappa \dd x \ ,
\end{equation}
and the needed first order variation of $\kappa$ is
\begin{equation}
    \delta \kappa = \frac{\phi''(x)}{(1 + (\L'(x))^2)^{3/2}} - \frac{3 \L''(x) \L'(x) \phi'(x)}{(1 + (\L'(x))^2)^{5/2}} \ .
\end{equation}
Denote $A(x) = \sqrt{1 + (\L'(x))^2}$. Then, the total first order index variation under such transfer is
\begin{equation}
    \delta \I_p^p[\kappa[\L(x)+\epsilon \phi(x)]] \approx p \int_0^1 (\kappa(x))^{p-1} \qty(\frac{\phi''(x)}{A(x)^3} - \frac{3 \L''(x) \L'(x) \phi'(x)}{A(x)^5}) \dd x \ .
\end{equation}
Fulfilling the transfer principle with such a perturbation reduces to finding the sign of this expression, which should be negative to reduce inequality. Integrate both expressions by parts to use the fact $\phi$ is positive. Boundary terms can be neglected by $\phi$'s boundary conditions \eqref{eq:phiboundaryconditions}, and we find
\begin{equation}
\begin{split}
    \delta \I_p^p(x) &= p \int_a^b (\kappa(x))^{p-1}\qty(\frac{\phi''(x)}{A(x)^3} - \frac{3 \L''(x) \L'(x) \phi'(x)}{A(x)^5}) \dd x \\
    &= p \int_a^b \qty[\qty(\frac{\kappa(x)^{p-1}}{A(x)^3})'' + \qty(\frac{3 \L''(x) \L'(x) \kappa(x)^{p-1}}{A(x)^5})'] \phi(x) \ \dd x \ .
\end{split}
\end{equation}
Since $\phi$ is positive, we find an explicit (albeit quite uninformative) criterion for such a transfer to reduce inequality for any given $p$:
\begin{equation}
    \qty(\frac{\kappa(x)^{p-1}}{A(x)^3})'' + \qty(\frac{3 \L''(x) \L'(x) \kappa(x)^{p-1}}{A(x)^5})' < 0 \ \forall \ x \in (a,b) \ .
\end{equation}
This is only the strict limit which is sufficient, but not necessary; in some parts of the curve the expression could be positive but still result in a net-negative change. As expected, the above bracketed criterion vanishes identically as $p = 1$, which is precisely consistent with $\I_1$ depending only on the endpoint values and satisfying Pigou--Dalton weakly.

Failure of the transfer principle for every $p > 1$ follows from the Lorenz dominance counterexample. In that case, $\L_2$ dominates $\L_1$, and can be reached from $\L_1$ by a sequence of progressive Pigou--Dalton transfers. Since $\I_p(\L_2) > \I_p(\L_1)$ for $p > p^* \approx 1.87$, at least one of these transfers strictly increases inequality. As discussed in the example, the smoothing parameter $\epsilon$ can be chosen such that the dominating curve is measured to be more unequal for any $p > 1$, and so the transfer principle can also be violated for any $p > 1$.

\section{Comparison with Atkinson- and Amato--Kakwani-indices}
\label{app:comparisonwithatkinsonandamatokakwaniindices}

Let us give a quick comparison of the $\I_p$-family with the classical index family of Atkinson \cite{Atkinson1970} and the Amato--Kakwani index \cite{Amato1968, Kakwani1980, Arnold2012} in their continuous formulations. 

The Atkinson's index family, in its continuous formulation is \cite{cowell2011measuring}
\begin{equation}
    A_\epsilon = 
    \begin{cases}
        & 1- \qty[\int_0^\infty \qty(\frac{y}{\mu})^{1-\epsilon}f(y) \ \dd y]^{\frac{1}{1 -\epsilon}} \ , \qquad \epsilon \neq 1 \\
        &1 - \exp \qty[\int_0^\infty \log \qty(\frac{y}{\mu}) f(y) \ \dd y] \ , \qquad \epsilon = 1 \ ,
    \end{cases}
\end{equation}
where $y$ is income, $f(y)$ is the income distribution, and $\mu$ is the mean income. The second form is a natural limiting case of the first. 

Let us focus on the first case and represent the index with the Lorenz curve. A direct change of variables for any probability density weighed integral,
\begin{equation}
    x = F(y) \ , \qquad y = \Q(x) \ , \qquad \dd x = f(y) \dd y \ ,
\end{equation}
allows writing an expectation value with the quantile function as
\begin{equation}
    \mathbb{E}[g(Y)] = \int_0^\infty g(y) f(y) \dd y = \int_0^1 g(Q(x)) \dd x \ .
\end{equation}
Hence, with $\L'(x) = \frac{\Q(x)}{\mu}$ we may write 
\begin{equation}
    \int_0^\infty y^{1-\epsilon} f(y) \dd y = \mu^{1-\epsilon} \int_0^1 (\L'(x))^{1-\epsilon} \dd x \ ,
\end{equation}
and the full $A_\epsilon$ becomes
\begin{equation}
    A_\epsilon = 1 - \qty(\int_0^1 (\L'(x))^{1-\epsilon} \dd x )^{\frac{1}{1 - \epsilon}} \ .
\end{equation}
Similarity with the curvature-based $\I_p$ is immediate. Strictly speaking, the inequality-aversion parameter $\epsilon$ ranges between 0 and $\infty$, but the second part of the right-hand side of the above is a norm only if $p = 1 - \epsilon \geq 1$, that is, if $\epsilon = 0$ exactly. For $0 < \epsilon < 1$, it is a quasi-norm, $\epsilon = 1$ a geometric mean, and $\epsilon > 1$ a power mean of negative order. In any case, the structural connection is clear. Abusing notation, one may denote
\begin{equation}
    A_\epsilon = 1 - ||\L'(x)||_{1 - \epsilon} \ .
\end{equation}
Another index to compare with is the one originally constructed by Amato and later popularized by Kakwani. It is based on the length of the Lorenz curve,
\begin{equation}
    \ell = \int_0^1 \sqrt{1 + (\L'(x))^2} \dd  x \ .
\end{equation}
The index itself is defined as the normalized version of the length, 
\begin{equation}
    K = \frac{\ell - \sqrt{2}}{2 - \sqrt{2}} \ .
\end{equation}
Comparing these two with the $\I_p$-family, we note important similarities: the stratification-aversion parameter is highly analogous and naturally inspired by the Atkinsonian inequality-aversion parameter. However, as we discussed, the range of possibilities for $\epsilon$ is not strictly a mathematical norm, but a generalized power mean of the slope of the Lorenz curve. In our case the stratification-aversion parameter ranges also between 0 and $\infty$, each value corresponding to a norm. As such, the theory of $L^p$-norms is directly applicable to the study of $\I_p$.

Given the shared $L^p$/power-mean structure, one may ask whether the $\I_p$-family is merely a reparametrization of the Atkinson family, that is, whether $\I_p = \Phi(A_{\epsilon(p)})$ for some monotone $\Phi$ and some assignment $p \mapsto \epsilon(p)$. It is not, and the structural reason is immediately visible above: for fixed $\epsilon$, the Atkinson index is a functional of the first derivative of the Lorenz curve alone. Fixing the mean and the value of $A_\epsilon$ leaves freedom in the second derivative, which the curvature family resolves. 

The same observation of $\L''(x)$ not being resolved applies verbatim to the Amato--Kakwani index, which is the $\sqrt{1 + (\L')^2}$-mean of the slope. The Amato--Kakwani index is especially comparable to $\I_p$ due to the principle governing the functional form of the integral over $\L'(x)$; it is specified directly by the study of the length of the curve instead of being a free choice. 

Thus, both of the comparison indices measure an averaged property of the first derivative of the Lorenz curve, and are sensitive to the changes in its slope. At the same time neither of them is sensitive to the second derivative of the Lorenz curve, which we have argued to measure social stratification and the existence of class boundaries. Studying curvature naturally includes the second derivative of $\L(x)$ into the index, with a similar single-parameter approach as the Atkinson index and a similar principle of construction as the Amato--Kakwani index.

\end{appendix}

\bibliography{bibliography.bib}

@Book{DeVore1993,
author={DeVore, Ronald A.
and Lorentz, G. G.},
title={Constructive approximation},
series={Grundlehren der mathematischen Wissenschaften: 303},
year={1993},
publisher={Springer-Verlag},
isbn={3540506276}
}

@book{cowell2011measuring,
  title={Measuring Inequality},
  author={Cowell, F.},
  isbn={9780191625121},
  series={London School of Economics Perspectives in Economic Analysis},
  url={https://books.google.fi/books?id=UoWV3FjS0MgC},
  year={2011},
  publisher={OUP Oxford}
}

@misc{WorldBankPIP,
  author       = {{World Bank}},
  title        = {{Poverty and Inequality Platform}},
  year         = {2026},
  howpublished = {World Bank Group. \url{https://pip.worldbank.org/}},
  note         = {Accessed 2026-07-12}
}

@book{OxfordHandbook,
    editor = {Salverda, Wiemer and Nolan, Brian and Smeeding, Timothy M.},
    title = {The Oxford Handbook of Economic Inequality},
    publisher = {Oxford University Press},
    year = {2011},
    month = {02},
    isbn = {9780199606061},
    doi = {10.1093/oxfordhb/9780199606061.001.0001},
    url = {https://doi.org/10.1093/oxfordhb/9780199606061.001.0001},
}

@article{Lorenz1905,
 ISSN = {15225437},
 URL = {http://www.jstor.org/stable/2276207},
 author = {M. O. Lorenz},
 journal = {Publications of the American Statistical Association},
 number = {70},
 pages = {209--219},
 publisher = {[American Statistical Association, Taylor & Francis, Ltd.]},
 title = {{Methods of Measuring the Concentration of Wealth}},
 urldate = {2026-07-08},
 volume = {9},
 year = {1905}
}

@book{Gini1912book,
  title={Variabilit{\`a} e mutabilit{\`a}: contributo allo studio delle distribuzioni e delle relazioni statistiche. [Fasc. I.]},
  author={Gini, C.},
  series={Studi economico-giuridici pubblicati per cura della facolt{\`a} di Giurisprudenza della R. Universit{\`a} di Cagliari},
  url={https://books.google.fi/books?id=fqjaBPMxB9kC},
  year={1912},
  publisher={Tipogr. di P. Cuppini}
}

@article{Gini1921paper,
 ISSN = {00130133, 14680297},
 URL = {http://www.jstor.org/stable/2223319},
 author = {Corrado Gini},
 journal = {The Economic Journal},
 number = {121},
 pages = {124--126},
 publisher = {[Royal Economic Society, Wiley]},
 title = {Measurement of Inequality of Incomes},
 urldate = {2026-07-08},
 volume = {31},
 year = {1921}
}

@Article{Atkinson1970,
author={Atkinson, Anthony B.},
title={On the measurement of inequality},
journal={Journal of Economic Theory},
year={1970},
month={Sep},
day={01},
volume={2},
number={3},
pages={244-263},
issn={0022-0531},
url={https://www.sciencedirect.com/science/article/pii/0022053170900396}
}

@book{Amato1968,
  title={Metodologia statistica strutturale},
  author={Amato, Vittorio},
  year={1968},
  publisher={F. Cacucci}
}

@Book{Kakwani1980,
author={Kakwani, Nanak.
and World Bank},
title={Income inequality and poverty : methods of estimation and policy applications },
series={A World Bank research publication},
year={1980},
publisher={Published for the World Bank by Oxford University Press},
address={New York},
isbn={0195202279}
}

@Article{Arnold2012,
author={Arnold, Barry C.},
title={{On the Amato inequality index}},
journal={Statistics {\&} Probability Letters},
year={2012},
month={Aug},
day={01},
volume={82},
number={8},
pages={1504-1506},
issn={0167-7152},
url={https://www.sciencedirect.com/science/article/pii/S0167715212001630}
}

@book{Theil1967,
  title={Economics and Information Theory},
  author={Theil, H.},
  isbn={9780444102829},
  lccn={67004596},
  series={Studies in mathematical and managerial economics},
  url={https://books.google.fi/books?id=VVNVAAAAMAAJ},
  year={1967},
  publisher={North-Holland Publishing Company}
}

@misc{WorldBankPIPData,
  author       = {{World Bank}},
  title        = {{Poverty and Inequality Platform (PIP): Percentiles. world\_100bin\_revised.csv}},
  year         = {2026},
  howpublished = {\url{https://datacatalog.worldbank.org/search/dataset/0063646/poverty-and-inequality-platform-pip-percentiles}},
  note         = {Accessed: 2026-07-12}
}

@Article{Savitzky1964,
author={Savitzky, Abraham.
and Golay, M. J. E.},
title={{Smoothing and Differentiation of Data by Simplified Least Squares Procedures}},
journal={Analytical Chemistry},
year={1964},
month={Jul},
day={01},
publisher={American Chemical Society},
volume={36},
number={8},
pages={1627-1639},
issn={0003-2700},
doi={10.1021/ac60214a047},
url={https://doi.org/10.1021/ac60214a047}
}

@article{ChauNan1982,
 ISSN = {00307653, 14643812},
 URL = {http://www.jstor.org/stable/2662589},
 author = {Chau-Nan Chen and Tien-Wang Tsaur and Tong-Shieng Rhai},
 journal = {Oxford Economic Papers},
 number = {3},
 pages = {473--478},
 publisher = {Oxford University Press},
 title = {{The Gini Coefficient and Negative Income}},
 urldate = {2026-07-08},
 volume = {34},
 year = {1982}
}

@article{Dalton1920,
 ISSN = {00130133, 14680297},
 URL = {http://www.jstor.org/stable/2223525},
 author = {Hugh Dalton},
 journal = {The Economic Journal},
 number = {119},
 pages = {348--361},
 publisher = {[Royal Economic Society, Wiley]},
 title = {{The Measurement of the Inequality of Incomes}},
 urldate = {2026-07-08},
 volume = {30},
 year = {1920}
}

@book{IncomeDistributionHandbook2014,
  title={Handbook of Income Distribution},
  editor={Atkinson, A.B. and Bourguignon, F.},
  isbn={9780444594761},
  series={Handbook of Income Distribution},
  url={https://books.google.fi/books?id=qwR-BAAAQBAJ},
  year={2014},
  publisher={North Holland}
}

@article{Esteban1994,
 ISSN = {00129682, 14680262},
 URL = {http://www.jstor.org/stable/2951734},
 author = {Joan-María Esteban and Debraj Ray},
 journal = {Econometrica},
 number = {4},
 pages = {819--851},
 publisher = {[Wiley, Econometric Society]},
 title = {{On the Measurement of Polarization}},
 urldate = {2026-07-08},
 volume = {62},
 year = {1994}
}

@article{Esteban2004,
 ISSN = {00129682, 14680262},
 URL = {http://www.jstor.org/stable/3598766},
 author = {Jean-Yves Duclos and Joan Esteban and Debraj Ray},
 journal = {Econometrica},
 number = {6},
 pages = {1737--1772},
 publisher = {[Wiley, Econometric Society]},
 title = {{Polarization: Concepts, Measurement, Estimation}},
 urldate = {2026-07-08},
 volume = {72},
 year = {2004}
}

@article{Wolfson1994,
 ISSN = {00028282},
 URL = {http://www.jstor.org/stable/2117858},
 author = {Michael C. Wolfson},
 journal = {The American Economic Review},
 number = {2},
 pages = {353--358},
 publisher = {American Economic Association},
 title = {{When Inequalities Diverge}},
 urldate = {2026-07-08},
 volume = {84},
 year = {1994}
}

@article{Wolfson2010,
  title = {{Polarization and the decline of the middle class: Canada and the {U.S.}}},
  author = {Foster, James E. and Wolfson, Michael C.},
  journal = {The Journal of Economic Inequality},
  year = {2010},
  volume = {8},
  number = {2},
  pages = {247--273},
  doi = {10.1007/s10888-009-9122-7},
  url = {https://link.springer.com/article/10.1007/s10888-009-9122-7}
}

@article{Yitzhaki1991,
author = {Yitzhaki, Shlomo and Lerman, Robert I.},
title = {{Income Stratification And Income Inequality}},
journal = {Review of Income and Wealth},
volume = {37},
number = {3},
pages = {313-329},
doi = {https://doi.org/10.1111/j.1475-4991.1991.tb00374.x},
url = {https://onlinelibrary.wiley.com/doi/abs/10.1111/j.1475-4991.1991.tb00374.x},
year = {1991}
}

@Article{Donaldson1980,
author={Donaldson, David
and Weymark, John A.},
title={A single-parameter generalization of the Gini indices of inequality},
journal={Journal of Economic Theory},
year={1980},
month={Feb},
day={01},
volume={22},
number={1},
pages={67-86},
issn={0022-0531},
url={https://www.sciencedirect.com/science/article/pii/0022053180900654}
}

@article{Yitzhaki1983,
 ISSN = {00206598, 14682354},
 URL = {http://www.jstor.org/stable/2648789},
 author = {Shlomo Yitzhaki},
 journal = {International Economic Review},
 number = {3},
 pages = {617--628},
 publisher = {[Economics Department of the University of Pennsylvania, Wiley, Institute of Social and Economic Research, Osaka University]},
 title = {{On an Extension of the Gini Inequality Index}},
 urldate = {2026-07-12},
 volume = {24},
 year = {1983}
}

@article{Gastwirth1972,
 ISSN = {00346535, 15309142},
 URL = {http://www.jstor.org/stable/1937992},
 author = {Joseph L. Gastwirth},
 journal = {The Review of Economics and Statistics},
 number = {3},
 pages = {306--316},
 publisher = {The MIT Press},
 title = {{The Estimation of the Lorenz Curve and Gini Index}},
 urldate = {2026-07-12},
 volume = {54},
 year = {1972}
}

\end{document}